\documentclass{article}
\usepackage{spconf}
\usepackage{amsmath}
\usepackage{graphicx}
\usepackage{booktabs}
\usepackage{threeparttable}
\usepackage{appendix}
\usepackage{algorithm}
\usepackage{algorithmic}
\usepackage{makecell}
\usepackage{multirow}
\usepackage{amsfonts}

\newcommand{\figref}{Fig.~\ref}
\newcommand{\tabref}{Table~\ref}
\newcommand{\neqref}{Eq.~\eqref}
\newcommand{\secref}{Section~\ref}

\title{Efficient Neural Architecture Search for End-to-end Speech Recognition via Straight-Through Gradients}
\name{Huahuan Zheng, Keyu An, Zhijian Ou\thanks{This work is supported by NSFC 61976122.}}
\address{Speech Processing and Machine Intelligence (SPMI) Lab, Tsinghua University, China. \\
         zhh20@mails.tsinghua.edu.cn, aky19@mails.tsinghua.edu.cn, ozj@tsinghua.edu.cn}
\begin{document}
\ninept
\setlength{\textfloatsep}{4pt}
\setlength{\abovedisplayskip}{3pt}
\setlength{\belowdisplayskip}{3pt}
\maketitle
\begin{abstract}
    Neural Architecture Search (NAS), the process of automating architecture engineering, is an appealing next step to advancing end-to-end Automatic Speech Recognition (ASR), replacing expert-designed networks with learned, task-specific architectures.
    In contrast to early computational-demanding NAS methods, recent gradient-based NAS methods, e.g., DARTS (Differentiable ARchiTecture Search), SNAS (Stochastic NAS) and ProxylessNAS, significantly improve the NAS efficiency.
    In this paper, we make two contributions.
    First, we rigorously develop an efficient NAS method via Straight-Through (ST) gradients, called ST-NAS.
    Basically, ST-NAS uses the loss from SNAS but uses ST to back-propagate gradients through discrete variables to optimize the loss, which is not revealed in ProxylessNAS.
    Using ST gradients to support sub-graph sampling is a core element to achieve efficient NAS beyond DARTS and SNAS.
    Second, we successfully apply ST-NAS to end-to-end ASR.
    Experiments over the widely benchmarked 80-hour WSJ and 300-hour Switchboard datasets show that the ST-NAS induced architectures significantly outperform the human-designed architecture across the two datasets.
    Strengths of ST-NAS such as architecture transferability and low computation cost in memory and time are also reported.
\end{abstract}
\section{Introduction}
\label{sec:intro}

Building Automatic Speech Recognition (ASR) systems historically was an expertise-intensive task and involved a complex pipeline \cite{soltau2005ibm, dahl2012context}, which consists of phonetic decision trees and multiple stages of alignments and model updating.
Recently, there are increasing interests in developing end-to-end ASR systems \cite{graves2006connectionist, graves2012sequence, chorowski2014end, xiang2019crf} to reduce expert efforts and simplify the system.
The progress largely relies on utilizing deep neural networks (DNNs) of various architectures, which is generally known as deep learning.
The success of deep learning is largely due to its automation of the feature engineering process: hierarchical feature extractors are automatically learned from data rather than manually designed.
This success has been accompanied, however, by a rising demand for architecture engineering.

Various neural architectures, e.g., 2D Convolutional Neural Networks (CNNs) \cite{simonyan2014very}, which is also known as (a.k.a.) VGG-Net, 1D dilated CNN \cite{peddinti2015time} (a.k.a. TDNN), ResNet \cite{he2016deep} and so on, are manually designed by experts through intuitions plus laborious trial and error experiments.
Hyper-parameters for an architecture (e.g., kernel size, stride, and dilation of CNN) are set empirically, which may not be optimal for the specific task at hand, since naive grid-search is highly expensive.
Neural Architecture Search (NAS) \cite{zoph2016neural}, the process of automating architecture engineering, is thus an appealing next step to advancing end-to-end ASR.

Early NAS methods are computationally demanding despite their remarkable performance. For example, it takes 2000 GPU days of reinforcement learning and 3150 GPU days of evolution to obtain a state-of-the-art (SOTA) architecture for image classification over CIFAR-10 \cite{zoph2018learning} and ImageNet \cite{real2019regularized} respectively .
Some recent NAS methods, e.g., DARTS (Differentiable ARchiTecture Search) \cite{liu2018darts}, SNAS (Stochastic NAS) \cite{xie2018snas} and ProxylessNAS \cite{cai2018proxelessnas}, significantly improve the NAS efficiency and can obtain SOTA architecture over CIFAR-10 within one GPU day.
Note that NAS methods roughly can be categorized according to three dimensions - search space, search method, and performance evaluation method \cite{elsken2019neural}.
Some common features shared by these efficient NAS methods are: representing the search space as a weighted directed acyclic graph (DAG)\footnote{When used to represent the entire architecture, the DAG is often called the super-network \cite{veniat2018learning,wu2019fbnet} or the over-parameterized network \cite{cai2018proxelessnas}.
    When the search space is defined over smaller building blocks \cite{pham2018efficient, liu2018darts}, the DAG is called the cell block or the cell, that could be stacked in some way via a preset or learned meta-architecture to form the entire architecture.}, using gradient-based search methods to learn the edge weights, and weight sharing in performance evaluation of candidate architectures (i.e., sub-graphs of the DAG).
The final architecture is derived from the learned edge weights.
Notably, DARTS uses the Softmax trick to relax the search space to be continuous and performs gradient search over the whole super-network, while SNAS uses the Gumbel-Softmax trick \cite{Jang2016Categorical}.
These are less efficient in both memory and computation than ProxylessNAS, which uses discrete search spaces and does sub-graph sampling.
To back-propagate gradients through the discrete variables, which index the sampled edges, ProxylessNAS uses an ad-hoc trick, analogous to BinaryConnect \cite{courbariaux2015binaryconnect}.

In this paper, we make two contributions. First, we observe that ProxylessNAS essentially uses the Straight-Through (ST) estimator \cite{bengio2013estimating} for gradient approximation, which is missed in the ProxylessNAS paper.
To back-propagate gradients through discrete variables, the basic idea of ST is that the sampled discrete index is used for forward computation, and the continuous Softmax probability of the sampled index is used for backward gradient calculation.
Using ST gradients to support sub-graph sampling is a core element to achieve efficient NAS beyond DARTS and SNAS.
Based on ProxylessNAS and also the above observation, we develop an efficient NAS method via ST gradients, called ST-NAS.
In contrast to ProxylessNAS whose NAS objective definition is not clearly shown, ST-NAS uses the NAS objective definition from SNAS and is more rigorous.
Compared to SNAS, ST-NAS does not use the Gumbel-Softmax trick and uses the more efficient ST gradients.
The development of the ST-NAS method is the first contribution of this paper.

Second, we apply ST-NAS to end-to-end ASR.
Compared to the application of NAS techniques in computer vision tasks, there are limited studies in applying NAS to speech recognition tasks.
In \cite{veniat2019stochastic}, NAS via policy gradient based reinforcement learning \cite{veniat2018learning} is applied to keyword spotting.
In \cite{mazzawi2019improving}, evolution-based NAS is applied to keyword spotting and language identification.
DARTS are used in \cite{ding2020autospeech} and \cite{chen2020darts} for speaker recognition and Connectionist Temporal Classification (CTC) based speech recognition respectively.
The NAS methods used in these previous studies are not as efficient as NAS via ST gradients. 
This work represents the first to introduce ST gradient based NAS into end-to-end speech recognition, which is the second contribution of this paper.

We evaluate the ST-NAS method in end-to-end ASR experiments over the widely benchmarked 80-hour WSJ and 300-hour Switchboard datasets and show that the ST-NAS induced architectures significantly outperform the human-designed architecture (TDNN-D in \cite{peddinti2018low}) across the two datasets.
Notably, our ST-NAS induced model obtains the lowest word error rate (WER) of 2.77\%/5.68\% on WSJ eval92/dev93 among all published end-to-end ASR results, to the best of our knowledge.
Our NAS implementation is based on the CAT toolkit \cite{an2020cat} - a PyTorch-based ASR toolkit, which offers two benefits. First, it enables us to seamlessly integrate the NAS code with the ASR code to flexibly use PyTorch functionalities. Second, the CAT toolkit supports the end-to-end CTC-CRF loss, which is defined by a CRF (conditional random field) with CTC topology and has been shown to perform significantly better than CTC \cite{xiang2019crf, an2020cat}.
Additionally, we show that the architectures learned by the CTC loss based ST-NAS are transferable to be retrained under the CTC-CRF loss, i.e., these architectures achieve close performance to the architectures searched under the CTC-CRF loss. This enables us to reduce the cost of running NAS to search the architecture, since CTC-CRF is somewhat more expensive than CTC.
We also show that the model transferred from the WSJ experiment performs close to the model searched over Switchboard and better than the comparably-sized TDNN-D.
We release the code\footnote{https://github.com/thu-spmi/ST-NAS} for reproducible NAS study, as it is found in \cite{Li2019Random} that reproducibility is crucial to foster NAS research.

\begin{figure}[t]
    \centering
    \includegraphics[width=8.5cm]{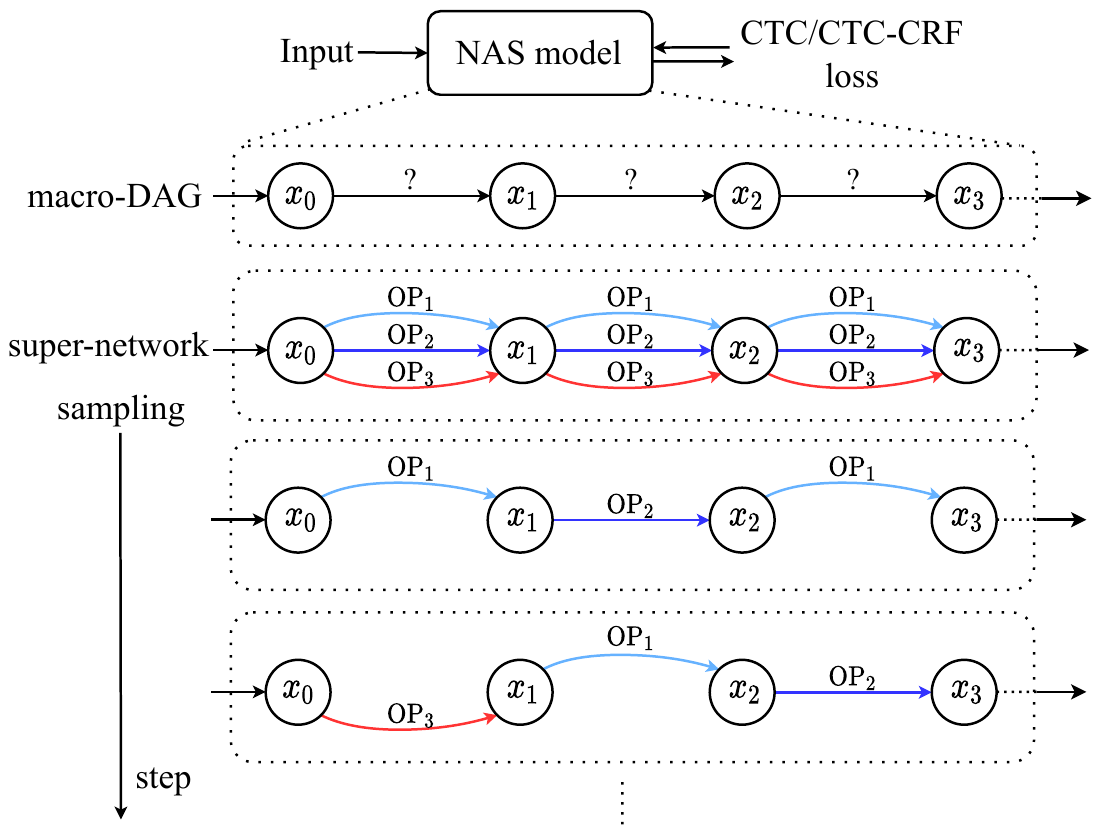}
    \vspace{-3mm}
    \caption{An overview of our ST-NAS method, which illustrates the concepts of macro-DAG, candidate operations (OP$_1$, OP$_2$, OP$_3$), super-network, and sub-graph sampling.
        Here we plot the macro-DAG used in our ASR experiments, which has a serial structure. More complicated macro-DAGs are possible in general.
    }
    \label{fig:toplevelsystem}
\end{figure}

\section{Background and Related Work}
\label{sec:relatedwork}

Recent gradient-based NAS methods such as DARTS \cite{liu2018darts}, SNAS \cite{xie2018snas} and ProxylessNAS \cite{cai2018proxelessnas} significantly improve over previous reinforcement learning and evolution based NAS methods with reduced computational cost.
In these methods, the search space is represented as a DAG, which consists of nodes (numbering by $0,1,\cdots,N$) and directed edges (pointing from lower-numbered nodes to higher-numbered). The $i$-th node, denoted by node$_i$, represents a latent representation $x_i$ (e.g., a feature map in CNNs). A directed edge leaving node$_i$ is associated with an operation that transforms $x_i$. An intermediate node (i.e., after the input node $x_0$) is computed as follows:
\begin{equation}
    x_j = \sum_{i \in \mathcal{A}_j}{\Omega_{ij}(x_i)}
    \label{eq:x_j}
\end{equation}
where $\mathcal{A}_j$ is the set of parent nodes of node$_j$, $\Omega_{ij}$ denotes the operation that connects node$_i$ to node$_j$ in the computation flow.
Suppose that $\Omega_{ij}$ can take from $K$ different candidate operations $\{o_{ij}^{(k)},k=1,\cdots,K\}$ (e.g., different convolutions), where each candidate operation is associated a weight $\alpha_{ij}^{(k)}$, called architecture weight.
It can be easily seen that by sampling one of the $K$ candidate operations for each connected pair of nodes, we obtain a candidate architecture.
The task of NAS therefore is reduced to learning the architecture weights. The final architecture is derived from the learned architecture weights, e.g., by selecting the largest weighted candidate operation for each connected pair of nodes $(i,j)$.
Different NAS methods mainly differ in how to learn the architecture weights $\alpha = \{\alpha_{ij}^{(k)}\}$, together with the operation parameters $\theta$, which are used to define the operations $\{o_{ij}^{(k)}\}$.

Note that it is often a practice to plot all the candidate operations $\{o_{ij}^{(k)},k=1,\cdots,K\}$ between each connected node$_i$ and node$_j$ in the DAG and call the resulting expanded DAG a super-network, e.g., as in \cite{liu2018darts,xie2018snas,cai2018proxelessnas} and also shown in \figref{fig:toplevelsystem}.
Then each edge in the super-network represents a candidate operation, and a candidate architecture corresponds to a sub-graph in the super-network.
Each connected pair of nodes together with the $K$ edges between them is called a searching block.
To be differentiated from the super-network, the initial DAG is referred to as the macro-DAG.

\subsection{DARTS}
\label{sec:relatedwork:DARTS}

Instead of searching over a discrete set of sub-graphs, DARTS relaxes the categorical choice of a particular operation $o_{ij}^{(k)}$ on edge $(i,j)$ to a mixture of all possible $K$ candidate operations, by defining the following mixed operation:
\begin{equation}
    \Omega_{ij}^{\text{DARTS}}(x_i) = \sum_{k=1}^K \pi_{ij}^{(k)} o_{ij}^{(k)}(x_i)
\end{equation}
where $\pi_{ij}=(\pi_{ij}^{(1)}, \cdots, \pi_{ij}^{(K)})$ denotes the mixing probabilities of the $K$ operations defined by a Softmax over the architecture weights:
\begin{equation}
    \pi_{ij}^{(k)}=\frac{\exp(\alpha_{ij}^{(k)})}{\sum_{k'=1}^{K} \exp(\alpha_{ij}^{(k')})}
    \label{eq:pi}
\end{equation}
Here we add DARTS as the superscript to denote the particular operation $\Omega_{ij}$ used in DARTS, which is plugged in \neqref{eq:x_j} to define the computation flow.
In this manner, DARTS relaxes the search space to be continuous and directly incorporates the architecture weights $\alpha$ into the super-network forward computation to define the loss $\mathcal{L}^{\text{DARTS}}(\alpha,\theta)$, together with the operation parameters $\theta$. Thus $\alpha$ and $\theta$ can be jointly trained via the standard backpropagation algorithm by optimizing $\mathcal{L}^{\text{DARTS}}(\alpha,\theta)$\footnote{In fact, $\mathcal{L}^{\text{DARTS}}_{\text{train}}(\alpha,\theta)$ with respect to (w.r.t.) $\theta$ and $\mathcal{L}^{\text{DARTS}}_{\text{val}}(\alpha,\theta)$ w.r.t.~$\alpha$ are alternately optimized, which are defined over training data and validation data respectively.}.

A limitation of DARTS is its expensive computation in both memory and time, which is around $K$ times of training a single model.
The mixed operation of DARTS requires to store the whole super-network in memory, and all edges are involved in the forward and backward computation.
\vspace{-2mm}

\subsection{SNAS}
\label{sec:relatedwork:SNAS}

Note that DARTS uses the super-network in NAS training, but after NAS training, uses the extracted sub-graph in inference. This incurs an inconsistency, which harms the performance.
SNAS uses the same super-network setup as in DARTS, but proposes to optimize the expected performance of all sub-graphs sampled with $p_\alpha(z)$:
\begin{equation}
    \mathcal{L}(\alpha,\theta) = \mathbb{E}_{z\sim p_{\alpha}(z)}[\mathcal{L}_\theta(z)]
    \label{eq:SNAS-loss}
\end{equation}
where $z = \{z_{ij}\}$ denotes the collection of the independent one-hot random variable vectors\footnote{As usual, categorical variables are encoded as $K$-dimensional one-hot vectors lying on the corners of the ($K-1$)-dimensional simplex.} for all pairs of connected nodes in the super-network, indexing, which edge is sampled. Thus a sample $z$ represents a sampled sub-graph, and $\mathcal{L}_\theta(z)$ denotes the loss evaluated under the sampled sub-graph $z$.
The one-hot vector $z_{ij}=(z_{ij}^{(1)}, \cdots, z_{ij}^{(K)})$ for connected node$_i$ and node$_j$ is assumed to follow the categorical distribution $Cat(\pi_{ij}^{(1)}, \cdots, \pi_{ij}^{(K)})$.

The loss $\mathcal{L}(\alpha,\theta)$ is not directly differentiable w.r.t.~the architecture weights $\alpha$, since we cannot pass the gradient through the discrete random variable $z_{ij}$ to $\alpha_{ij}=(\alpha_{ij}^{(1)}, \cdots, \alpha_{ij}^{(K)})$.
To sidestep this, SNAS relaxes the discrete one-hot variable $z_{ij}$ to be a continuous random variable computed by the Gumbel-Softmax (G-S) function \cite{Jang2016Categorical}:
\begin{equation}
    y_{ij}^{(k)}=\frac{\exp((\alpha_{ij}^{(k)}+g_{ij}^{(k)})/\tau)}{\sum_{k'=1}^{K} \exp((\alpha_{ij}^{(k')}+g_{ij}^{(k')})/\tau)}
\end{equation}
where $g_{ij}^{(k)}=-\log(-\log(u_{ij}^{(k)}))$, and $\{u_{ij}^{(k)}\}$ are independent and identical distributed (i.i.d.) samples from $u\sim \text{Uniform}(0,1)$. $\tau$ is the temperature, which is gradually annealed to be close to zero.

The soften one-hot variable $y_{ij}=(y_{ij}^{(1)}, \cdots, y_{ij}^{(K)})$ follows the G-S distribution.
It is shown in \cite{Jang2016Categorical} that as the temperature $\tau$ approaches $0$, $y_{ij}$ from the G-S distribution become one-hot, and the G-S distribution becomes identical to the categorical distribution $Cat(\pi_{ij}^{(1)}, \cdots, \pi_{ij}^{(K)})$. Thus, SNAS uses the following surrogate loss to approximate the loss defined in \neqref{eq:SNAS-loss}:
\begin{equation}
    \tilde{\mathcal{L}}(\alpha,\theta) = \mathbb{E}_{y\sim p_{\alpha}(y)}[\mathcal{L}_\theta(y)]
\end{equation}
which becomes directly differentiable w.r.t.~$\alpha$.
Here $\mathcal{L}_\theta(y)$ is defined by the computation flow still according to \neqref{eq:x_j} but with the following soften operation:
\begin{equation}
    \Omega_{ij}^{\text{SNAS}}(x_i) = \sum_{k=1}^K y_{ij}^{(k)} o_{ij}^{(k)}(x_i)
\end{equation}

It can be seen that the computation cost of SNAS in both memory and time is identical to that of DARTS.
The difference is that while DARTS uses the Softmax trick for continuous relaxation, SNAS uses the Gumbel-Softmax trick with annealed temperature. When $\tau$ approaches $0$, the objectives in NAS training and in model inference becomes consistent. The Gumbel-Softmax trick is also used in \cite{wu2019fbnet, dong2019searching} for NAS.

\subsection{ProxylessNAS}
\label{sec:relatedwork:pNAS}
Using the same super-network setup as in DARTS, ProxylessNAS proposes to use binary gates (essentially a one-hot vector) $z_{ij}$ for each edge to define the operation to reduce the memory footprint:
\begin{equation}
    \Omega_{ij}(x_i) = \sum_{k=1}^K z_{ij}^{(k)} o_{ij}^{(k)}(x_i),
    \label{eq:z-operation}
\end{equation}
Here we use the same $z_{ij}$ from SNAS as it carries the same meaning - indexing, which edge is sampled.
In this manner, the architecture weights $\alpha_{ij}$ are not directly involved in the computation flow as defined in \neqref{eq:x_j} and thereby we cannot directly calculate the gradient of $\alpha_{ij}$.
Motivated by BinaryConnect \cite{courbariaux2015binaryconnect}, ProxylessNAS proposes the following gradient approximation:
\begin{equation}
    \begin{aligned}
        \frac{\partial{\mathcal{L}}}{\partial{\alpha_{ij}^{(k)}}}
        = \sum_{k'=1}^K \frac{\partial{\mathcal{L}}}{\partial{\pi_{ij}^{(k')}}} \frac{\partial{\pi_{ij}^{(k')}}}{\partial{\alpha_{ij}^{(k)}}}
        \approx \sum_{k'=1}^K \frac{\partial{\mathcal{L}}}{\partial{z_{ij}^{(k')}}} \frac{\partial{\pi_{ij}^{(k')}}}{\partial{\alpha_{ij}^{(k)}}}
    \end{aligned}
    \label{eq:proxyless}
\end{equation}
By using the sampled sub-graph $z$, ProxylessNAS reduces the memory footprint and backward computation cost, compared to DARTS and SNAS.
We leave the detailed comparison of SNAS, ProxylessNAS and our method to \secref{sec:method}.

\section{method}
\label{sec:method}

Our NAS method is based on the following two observations from the review of existing gradient-based NAS methods.
First, ProxylessNAS is an interesting method, but the loss $\mathcal{L}$ used in ProxylessNAS is not explicitly shown in the original paper \cite{cai2018proxelessnas}. We observe that the ProxylessNAS loss $\mathcal{L}$ in \neqref{eq:proxyless} is in fact the loss $\mathcal{L}_\theta(z)$ as defined in SNAS.
Second, we observe that ProxylessNAS essentially uses the Straight-Through (ST) estimator \cite{bengio2013estimating} for gradient approximation - a simple yet effective technique to back-propagate gradients through discrete variables. This point is also missed in the ProxylessNAS paper.

In the following, we first introduce the ST gradient estimator, and then present our NAS method via ST gradients, called ST-NAS. Basically, ST-NAS uses the loss from SNAS but optimizes the loss using the ST gradients.

\subsection{Straight-Through gradient}
To optimize the loss $\mathcal{L}(\alpha,\theta)$, we sample a sub-graph, represented by $z$, according to the architecture weights $\alpha$.
Specifically, for each edge $(i,j)$, $z_{ij}$ is independently sampled from $Cat(\pi_{ij}^{(1)}, \cdots, \pi_{ij}^{(K)})$, which is defined by the architecture weights $\alpha_{ij}$ as in \neqref{eq:pi}.
Then the forward computation is conducted over the sampled sub-graph to obtain the loss $\mathcal{L}_\theta(z)$ as a Monte Carlo estimate of $\mathcal{L}(\alpha,\theta)$, by executing the operation as defined in \neqref{eq:z-operation}.
During the backward computation, we encounter the problem that we cannot pass the gradient through the one-hot sample $z_{ij}$ to $\alpha_{ij}$.

The basic idea of ST is that the sampled discrete $z_{ij}$ is used for forward computation, but the continuous probability $\pi_{ij}$ is used for backward gradient calculation, namely approximating $\partial{z_{ij}} \approx \partial{\pi_{ij}}$\footnote{Clearly, this is the approximation used by \neqref{eq:proxyless} in ProxylessNAS.}. Specifically, we compute the gradient w.r.t.~$\alpha_{ij}$ as follows:
\begin{equation}
    \begin{aligned}
        \frac{\partial{\mathcal{L}_\theta(z)}}{\partial{\alpha_{ij}^{(k)}}}
         & =\frac{\partial{\mathcal{L}_\theta(z)}}{\partial{x_j}} \frac{\partial{x_j}}{\partial{\alpha_{ij}^{(k)}}} =
        \frac{\partial{\mathcal{L}_\theta(z)}}{\partial{x_j}} \sum_{i\in\mathcal{A}_j} \frac{\partial{\Omega_{ij}(x_i)}}{\partial{\alpha_{ij}^{(k)}}}                                                           \\
        \frac{\partial{\Omega_{ij}(x_i)}}{\partial{\alpha_{ij}^{(k)}}}
         & =\sum_{k'=1}^K \frac{\partial{z_{ij}^{(k')}}}{\partial{\alpha_{ij}^{(k)}}} o_{ij}^{(k')}(x_i) \approx \sum_{k'=1}^K \frac{\partial{\pi_{ij}^{(k')}}}{\partial{\alpha_{ij}^{(k)}}} o_{ij}^{(k')}(x_i)
    \end{aligned}
    \label{eq:ST-grad}
\end{equation}
\begin{figure}[t]
    \begin{minipage}[b]{0.32\linewidth}
        \centering
        \includegraphics[width=2.8cm]{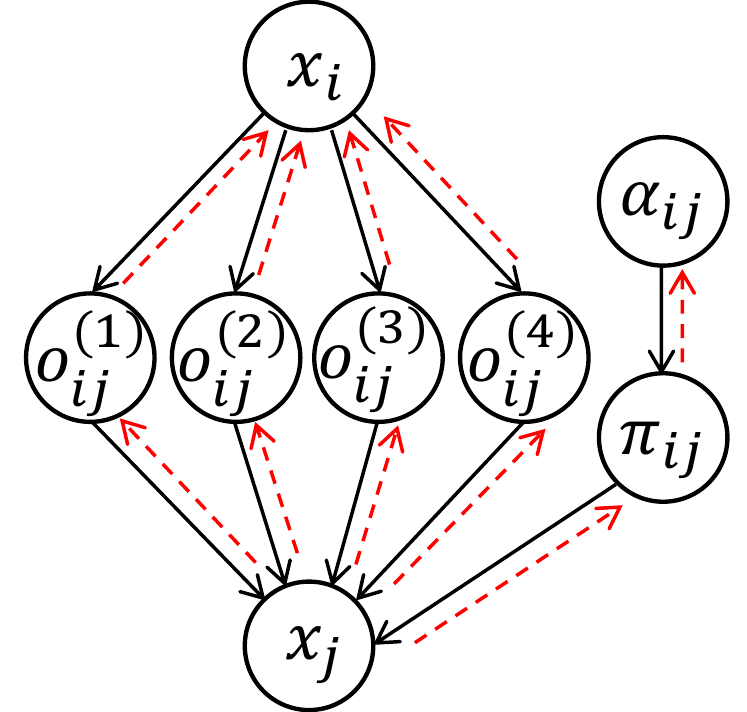}
        \centerline{(a) DARTS}\medskip
    \end{minipage}
    \begin{minipage}[b]{0.32\linewidth}
        \centering
        \includegraphics[width=2.8cm]{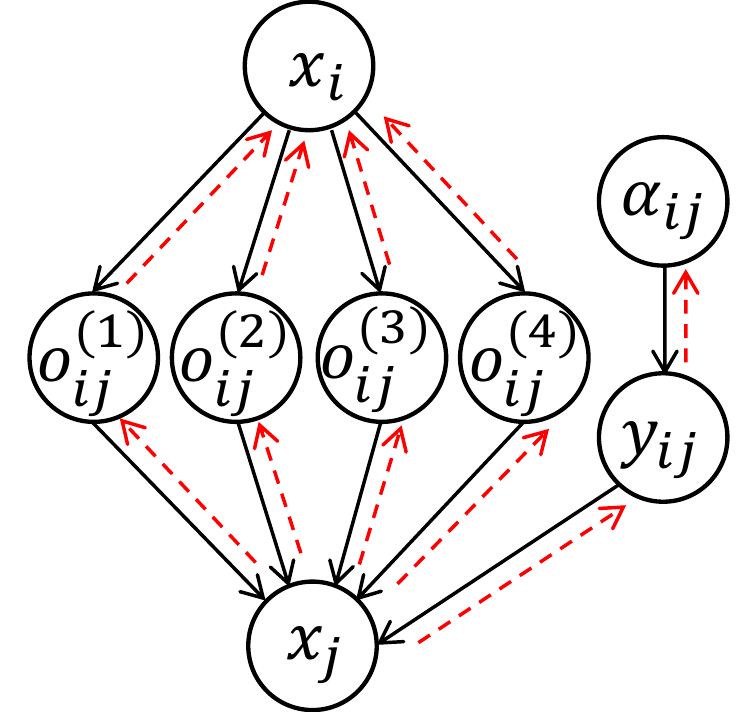}
        \centerline{(b) SNAS}\medskip
    \end{minipage}
    \begin{minipage}[b]{0.32\linewidth}
        \centering
        \includegraphics[width=2.8cm]{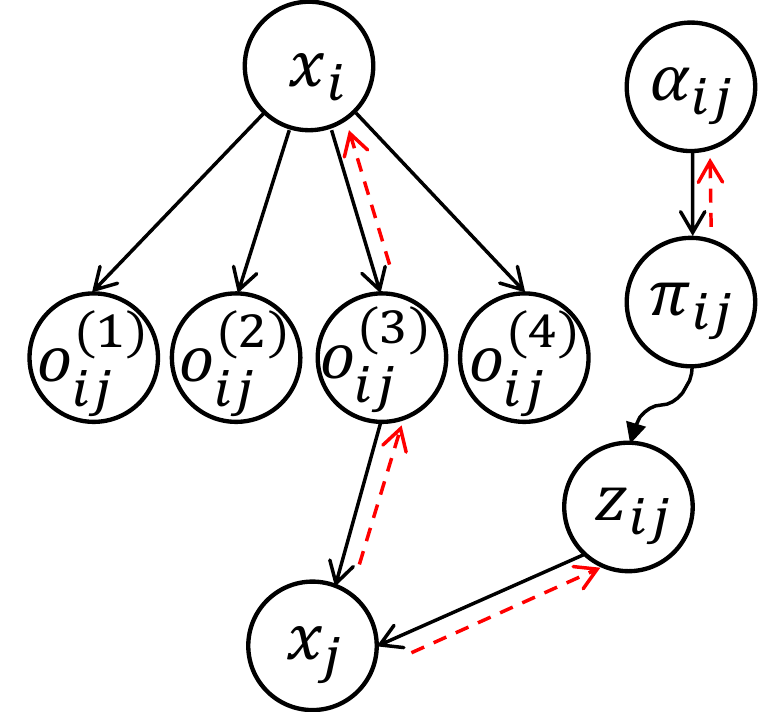}
        \centerline{(c) ST-NAS}\medskip
    \end{minipage}
    \vspace{-4mm}
    \caption{Computation flow of different NAS methods locally between connected node$_i$ and node$_j$ when $K=4$. Solid and dashed lines denote the forward and backward computations respectively. (a) For DARTS and SNAS, the forward is fully continuous and the backward is fully differentiable. (b) For ProxylessNAS and ST-NAS, the forward from $\pi_{ij}$ to $z_{ij}$ involves sampling and the backward uses the ST gradient to flow from $z_{ij}$ to $\alpha_{ij}$.}
    \label{fig:compute-flow}
\end{figure}

An illustration of the forward and backward computation performed locally between connected node$_i$ and node$_j$ is shown in \figref{fig:compute-flow}.
We compare different gradient-based NAS methods in \tabref{table:tab1}.
Compared to training a single model, both DARTS and SNAS are around $K$ times more expensive in both memory and time.
Note that as shown in \neqref{eq:ST-grad}, the gradient w.r.t.~$\alpha_{ij}$ involves all the $K$ features $\{o_{ij}^{(k')}(x_i), k'=1,\cdots,K\}$.
Regarding this, the backward computation complexity in ProxylessNAS and ST-NAS can be reduced to be $O(1)$, but the forward computation complexity is still $O(K)$.
As shown in \tabref{table:tab1}, the memory cost in ProxylessNAS and ST-NAS is far less than $K C_1$, where $C_1$ denotes the memory size for training a single model.

\begin{table}[t]
    \centering
    \vspace{-4mm}
    \caption{Comparison of different gradient-based NAS methods.\tnote{1}}
    \vspace{-2mm}
    \resizebox{0.48\textwidth}{!}{
        \begin{threeparttable}
            \label{table:tab1}
            \begin{tabular}{ccccc}
                \toprule
                Methods   & Loss                                         & $\alpha$ gradient & Memory          & \makecell{Backward \\computation} \\
                \midrule
                DARTS     & $\mathcal{L}^{\text{DARTS}}(\alpha, \theta)$ & continuous        & $K C_1$         & $O(K)$             \\
                SNAS      & $\tilde{\mathcal{L}}(\alpha, \theta)$        & continuous        & $K C_1$         & $O(K)$             \\
                Proxyless & $\mathcal{L}_\theta(z)$                      & ST                & $C_1+(K-1) C_2$ & $O(1)$             \\
                ST-NAS    & $\mathcal{L}(\alpha, \theta)$                & ST                & $C_1+(K-1) C_2$ & $O(1)$             \\
                \bottomrule
            \end{tabular}
            \begin{tablenotes}
                \footnotesize
                \item[1] Computation costs are estimated relative to training a single model. $K$ denotes the number of possible operations for each connected pair of nodes. The forward computation complexity of all methods is $O(K)$.
                $C_1$ denotes the memory size for training a single model.
                $C_2$ denotes the average memory size for storing the output features for all connected pairs of nodes in a sub-graph.
                Usually we have $C_2\ll C_1$ (see numerics in \secref{sec:exp:resource}).
            \end{tablenotes}
        \end{threeparttable}}
\end{table}
\vspace{-2mm}
\subsection{The NAS procedure}
\label{sec:method:nasprocedure}
Here we present the whole NAS procedure, which uses the ST gradients and is illustrated in \figref{fig:toplevelsystem}.
First, we need to choose the macro-DAG and the set of possible candidate operations, which together define the super-network, as the search space.
The setup of macro-DAG, candidate operations and the super-network is flexible and depends on the target task.
One example setup for the end-to-end ASR task is described in \secref{sec:exp}.
Given this setup, we divide the NAS procedure into three stages: super-network initialization (also called warm-up), architecture search and retraining.

There are two sets of learnable parameters: the architecture weights and operation parameters, denoted by $\alpha$ and $\theta$ respectively.
We split the original dataset into a training set and a validation set.

\textbf{Super-network initialization.}
Note that the single set of operation parameters $\theta$ is shared for all sampled sub-graphs, and thus plays a critical role in later architecture search.
It is found in our experiments that an initialization stage to warm up $\theta$ is beneficial.
Basically, this initialization stage is similar to the following architecture search stage, except that only $\theta$ is trained.
Specifically, for each minibatch from the training data, we uniformly sample a sub-graph, update $\theta$ using the standard gradient descent.
After each training epoch, we evaluate the super-network over the validation data (which is to be detailed below) to monitor the convergence.
The effect of this initialization stage on the performance of the searched architecture is detailed in \secref{sec:method:effectSI}.

%

\textbf{Architecture search.}
After completing the super-network initialization, we hold $\theta$, reset the optimizer and run the architecture search stage, which involves a bi-level optimization problem \cite{Colson2007an}:
\begin{equation}
    \begin{aligned}
                        & \min_{\alpha} \mathcal{L}_{val}(\alpha, \theta^*(\alpha))                         \\
        \text{s.t.}\ \  & \theta^*(\alpha) = \mathop{\arg\min}_{\theta} \mathcal{L}_{train}(\alpha, \theta)
    \end{aligned}
\end{equation}
where $\mathcal{L}_{train}$ and $\mathcal{L}_{val}$ are the losses over the training and validation data respectively.
Solving the above bi-level optimization problem is difficult. In practice, updating $\alpha$ and $\theta$ alternately by stochastically optimizing $\mathcal{L}_{val}$ and  $\mathcal{L}_{train}$ over minibatches respectively is found to work reasonably well, as shown in many previous NAS methods and given in Algorithm \ref{alg:AS}.
Note that as the validation set is usually smaller than the
training set, we cyclically use the validation set in drawing validation minibatches in \emph{Step 1} of Algorithm \ref{alg:AS}.
For evaluating the super-network over the validation data to monitor convergence, we average the validation losses over the minibatches in the whole validation set (not cyclically). Specifically, for each validation minibatch, we sample a sub-graph and calculate the validation loss over the sampled sub-graph.
In this manner, we evaluate the expected performance of the current super-network.
\begin{algorithm}[t]
    \caption{Architecture Search}
    \label{alg:AS}
    \begin{algorithmic}
        \WHILE{\text{not converged}}
        \FOR{each training minibatch in an epoch}
        \STATE \emph{Step 1.} Freeze $\theta$, draw a validation minibatch, sample a sub-graph, run the forward computation\footnotemark over the super-network under the validation minibatch, and update $\alpha$ with the ST gradients;
        \STATE \emph{Step 2.} Freeze $\alpha$, sample a sub-graph, run the forward computation over the sampled sub-graph under the training minibatch, and update $\theta$ with the standard gradients;
        \ENDFOR
        \STATE Evaluate the super-network over the validation data to monitor convergence.
        \ENDWHILE
    \end{algorithmic}
\end{algorithm}
\footnotetext{In this forward computation, we calculate $\{o_{ij}^{(k')}(x_i), k'=1,\cdots,K\}$, but use $\Omega_{ij}(x_i)$ as defined in \neqref{eq:z-operation}.}

\textbf{Retraining.}
After the convergence of the architecture search, we derive a single model by selecting the top-$1$ edge between connected nodes and prune others from the super-network.
The derived single model is trained from scratch to yield the final model.

\vspace{-3mm}
\section{Experiments}
\label{sec:exp}
\vspace{-1mm}
Experiments are conducted on the 80-hour WSJ and the 300-hour Switchboard datasets. Input features are 40 dimension fbank with delta and delta-delta features (120 dimensions in total). The features are augmented by 3-fold speed perturbation, and are mean and variance normalized.
We use the CAT toolkit \cite{an2020cat} for all experiments.
\begin{figure*}[t]
    \begin{minipage}[t]{1.0\linewidth}
        \centering
        \includegraphics[width=17cm]{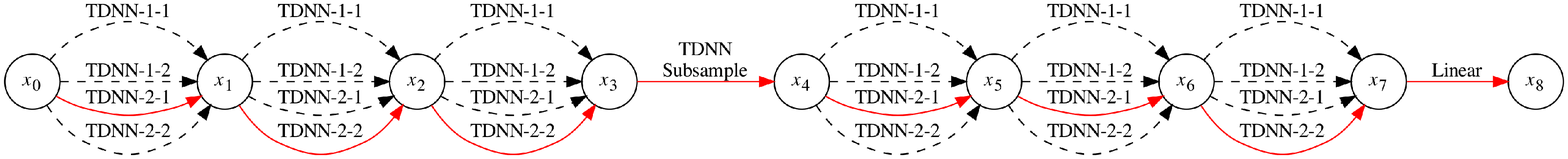}
        \centerline{(a)}\medskip
    \end{minipage}
    \begin{minipage}[b]{1.0\linewidth}
        \centering
        \includegraphics[width=17cm]{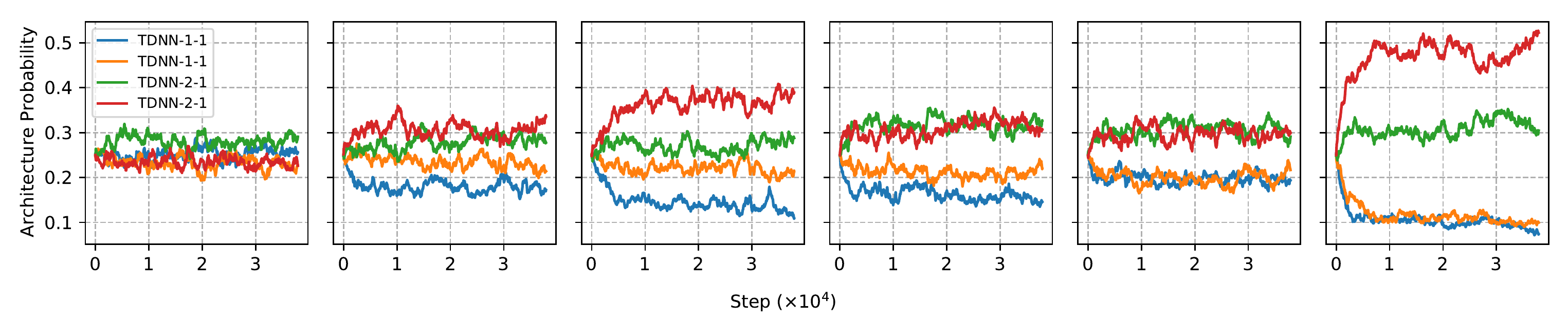}
        \centerline{(b)}\medskip
    \end{minipage}
    \vspace{-8mm}
    \caption{(a) shows the super-network for WSJ experiments. Labels of candidate operations over edges are formatted in ``-\{half of context\}-\{dilation\}''. In Switchboard experiments, we add extra ``TDNN-3-1'' and ``TDNN-3-2'' candidate operations to each searching blocks. The solid lines in (a) indicate one of the derived single model from the 5 runs of NAS on WSJ.
        (b) shows the evolution of architecture probabilities (i.e., $\pi_{ij}^{(k)}$) for the searching blocks in the NAS run that yields the derived single model in (a).}
    \label{fig:fig3}
    \vspace{-6mm}
\end{figure*}
\vspace{-2mm}
\subsection{NAS settings and hyper-parameters}
\label{sec:exp:settings}
We conduct NAS for the acoustic model.
Unless otherwise stated, we follow the basic settings in \cite{xiang2019crf}.
The denominator graph uses a phone-based 4-gram language model, and the language model in decoding is a word-based 4-gram language model.
For NAS settings, we design a super-network by serially connecting 3 searching blocks, 1 subsampling TDNN layer, 3 searching blocks, and 1 fully-connected layer, as shown in \figref{fig:fig3}(a). Although more complex super-networks can be designed, this super-network is inspired by the ``TDNN-D'' of \cite{peddinti2018low}. It can be easily seen that the ``TDNN-D'' is a sub-graph in the super-network, thus lies in our search space.

The subsampling layer is a TDNN layer with convolution kernel size 3, dilation 1, and stride 3, which leads to a subsampling of factor 3.
The candidate operations in each searching block are TDNNs with different configurations.
The number of hidden units in all TDNN layers is 640.
Convolution strides of candidate TDNNs are all set to 1.
Layer normalization and dropout with probability 0.5 are applied after each TDNN layer.

By default, during the warm-up and architecture search stages, the super-network is trained with CTC. This is called NAS with CTC.
The derived single model is then retrained with CTC-CRF.
We run the 3-stage NAS procedure as described in \secref{sec:method:nasprocedure} with the following hyper-parameters. We use the Adam optimizer (in PyTorch) with its default arguments unless otherwise stated. It can be seen that for simplicity and reproducibility, we only introduce a few extra hyper-parameters compared to training a single model.
\begin{itemize}
    \item Super-network initialization: 
          minibatch size is 128, learning rate is fixed to $10^{-3}$. Iterations stop if no improvement of the validation loss for 3 training epochs.
    \item Architecture search: 
          minibatch size is 64, learning rate decays from $10^{-3}$ to $10^{-4}$ with 0.1 decay rate if no improvement of the validation loss for 3 training epochs. Iterations stop if learning rate smaller than $10^{-4}$.
    \item Retraining: 
          minibatch size is 128, learning rate decays from $10^{-3}$ to $10^{-5}$ with 0.1 decay rate if no improvement of the validation loss for 1 training epoch. Iterations stop if learning rate smaller than $10^{-5}$.
\end{itemize}

\vspace{-2mm}
\subsection{WSJ}
\label{sec:exp:wsj}
\vspace{-1mm}
The candidate operations of WSJ experiments are shown in \figref{fig:fig3}(a), with which the search space contains 4096 sub-graphs in total.
Eval92 and dev93 sets are both for test and excluded from training set and validation set.
In warm-up and architecture search, the original training data are split into 90\%:10\% proportions for training and validation. For retraining, 5\% sentences of the original training set are for validation and the others for training.
The experiments run on 4 NVIDIA GTX1080 GPUs. To reduce the randomness, we conduct NAS with CTC for 5 times with different random seeds. The retraining uses the CTC-CRF loss with the CTC loss (weighted by 0.01). For comparison, NAS with fully CTC-CRF (namely the warm-up and architecture are trained with CTC-CRF) is conducted 3 times due to the expensive computation of CTC-CRF. Random search is conducted 5 times.
\begin{table}[t]
    \centering
    \vspace{-2mm}
    \caption{WERs on the 80-hour WSJ.}
    \resizebox{0.435\textwidth}{!}{
        \begin{threeparttable}
            \label{table:tab2}
            \begin{tabular}{cccc}
                \toprule
                \multicolumn{2}{c}{Methods}                                  & eval92                                 & dev93                                     \\
                \midrule
                \multicolumn{2}{c}{EE-Policy-CTC \cite{Zhou2018Improving}}   & 5.53                                   & 9.21                                      \\
                \multicolumn{2}{c}{SS-LF-MMI \cite{Hadian2018flatstart}}     & 3.0                                    & 6.0                                       \\
                \multicolumn{2}{c}{EE-LF-MMI \cite{Hadian2018e2e}}           & 3.0                                    & -                                         \\
                \multicolumn{2}{c}{FC-SR \cite{Zeghidour2018fully}\tnote{1}} & 3.5                                    & 6.8                                       \\
                \multicolumn{2}{c}{ESPRESSO \cite{Wang2019Espresso}}         & 3.4                                    & 5.9                                       \\
                \midrule
                \multicolumn{1}{c}{\multirow{2}{*}{CTC}}                     & BLSTM                                  & 4.93                & 8.57                \\
                ~
                ~                                                            & ST-NAS                                 & 4.72$\pm$0.03       & 8.82$\pm$0.07       \\
                \midrule
                \multicolumn{1}{c}{\multirow{5}{*}{CTC-CRF}}                 & BLSTM \cite{xiang2019crf}              & 3.79                & 6.23                \\
                ~                                                            & VGG-BLSTM \cite{an2020cat}             & 3.2                 & 5.7                 \\
                \cmidrule{2-4}
                ~                                                            & TDNN-D\tnote{2} \cite{peddinti2018low} & 2.91                & 6.24                \\
                ~                                                            & Random search\tnote{3}                 & 2.82$\pm$0.01       & 5.71$\pm$0.03       \\
                ~                                                            & ST-NAS                                 & {\bf 2.77}$\pm$0.00 & {\bf 5.68}$\pm$0.01 \\
                \midrule
                \multicolumn{2}{c}{ST-NAS with fully CTC-CRF}                & 2.81$\pm$0.01                          & 5.74$\pm$0.02                             \\
                \bottomrule
            \end{tabular}
            \begin{tablenotes}
                \footnotesize
                \item[1] FC-SR uses dev93 as validation set and eval92 for test.
                \item[2] Obtained based on our implementation of the ``TDNN-D'' in \cite{peddinti2018low}.
                \item[3] Random search is a competitive baseline for NAS \cite{Li2019Random} but still inferior to our ST-NAS.
            \end{tablenotes}
        \end{threeparttable}}
\end{table}

We compare our searched models to various human-designed DNN architectures in end-to-end ASR, trained with CTC, CTC-CRF or attention-based losses.
As shown in \tabref{table:tab2}, the model searched by ST-NAS with CTC and retrained with CTC-CRF achieves the lowest WER of 2.77\%/5.68\% on WSJ eval92/dev93, outperforming all other end-to-end ASR models.
This model obtains significant improvement over both BLSTM and VGG-BLSTM with CTC-CRF, with 26.9\% and 13.4\% relative WER reductions respectively on eval92.
On dev93, this model shows an 8.8\% relative improvement over BLSTM and is close to VGG-BLSTM.
Notably, the number of parameters in our searched models on average is around 11.9 million, 11.8\% less than BLSTM and 25.6\% less than VGG-BLSTM.

There are some ablation results in \tabref{table:tab2}.
First, for the models searched by NAS with CTC, retraining with CTC-CRF loss achieves improvement over retraining with CTC loss.
Second, compared to the models searched and retrained all with CTC-CRF (namely NAS with fully CTC-CRF), the models searched with CTC but retrained with CTC-CRF perform equally well.
In summary, these results show that the architectures searched by NAS with CTC are transferable to be retrained with CTC-CRF.
This enables us to reduce the cost of running NAS to search the architecture, since CTC-CRF is somewhat expensive than CTC.

\vspace{-3mm}
\subsection{Switchboard}
\vspace{-1mm}
In the Switchboard experiment, we add extra ``TDNN-3-1'' and ``TDNN-3-2'' candidate operations, in addition to those shown in \figref{fig:fig3}(a) for WSJ. The entire search space contains 46656 sub-graphs.
The retraining uses the CTC-CRF loss with the CTC loss (weighted by 0.1). In warm-up and architecture search, the original training data are split into 95\%:5\% proportions for training and validation.
For retraining, we take around 5 hours of speech as validation set and the rest for training, following the setting in CAT \cite{an2020cat}.
The Eval2000 data, including the Switchboard evaluation dataset (SW) and Callhome evaluation dataset (CH), are for testing.
We conduct a single run of ST-NAS on 4 NVIDIA P100 GPUs, and do not conduct multiple random searches due to time limitation.

As shown in \tabref{table:tab3}, our ST-NAS models outperform both the TDNN-D-Small and TDNN-D-Large.
Notably, the model transferred from the WSJ experiment performs close to the model searched over Switchboard; and compared to the comparably-sized TDNN-D-Large, it obtains 13.7\% and 8.6\% relative WER reductions on SW and CH respectively.

\begin{table}[t]
    \centering
    \caption{WERs on the 300-hour Switchboard\tnote{1}.}
    \vspace{-2mm}
    \resizebox{0.44\textwidth}{!}{
        \begin{threeparttable}
            \label{table:tab3}
            \begin{tabular}{ccccc}
                \toprule
                \multicolumn{2}{c}{Methods}                 & SW                            & CH   & Params          \\
                \midrule
                \multicolumn{2}{c}{TDNN-D-Small}            & 15.2                          & 26.8 & 7.64M           \\
                \multicolumn{2}{c}{TDNN-D-Large}            & 14.6                          & 25.5 & 11.85M          \\
                \midrule
                \multicolumn{1}{c}{\multirow{2}{*}{ST-NAS}} & Transferred from WSJ\tnote{2} & 12.5 & 23.2   & 11.89M \\
                ~                                           & Searched on Switchboard       & 12.6 & 23.2   & 15.98M \\
                \bottomrule
            \end{tabular}
            \begin{tablenotes}
                \footnotesize
                \item[1] All experiments are trained with CTC-CRF. TDNN-D-Small is with the hidden size of 640, which is the same as that of our searched models. TDNN-D-Large is with the hidden size of 800.
                \item[2] Randomly taken from one of the 5 runs of NAS with CTC over WSJ, and retrained on Switchboard.
            \end{tablenotes}
        \end{threeparttable}}
\end{table}

\vspace{-3mm}
\subsection{Complexity analysis}
\label{sec:exp:resource}
\vspace{-1mm}
To complement \tabref{table:tab1}, this section provides numerical complexity analysis in memory and time for running ST-NAS.
In the WSJ experiment, there are 6 searching blocks in our super-network and we run on 4 GPUs with data parallel in PyTorch.
Training a single model requires around $C_1=3.5\text{GB}$ memory on each GPU.
The quantity $C_2$ in \tabref{table:tab1} can be calculated as follows:
\begin{equation}
    \begin{aligned}
        C_2 & = 6 \times \text{MinibatchSize} \times \text{SequenceLen}                          \\
            & \times \text{HiddenUnitsNum}\times 4\text{~Byte} \div \text{GPUNum}                \\
            & =6\times 64 \times 850 \times 640 \times 4\text{~Byte} \div 4 \approx 209\text{MB}
    \end{aligned}
\end{equation}
which is far less than $C_1$. $\text{SequenceLen}$ denotes the average length of sequences, which is around 850.

For time complexity, warm-up and architecture search are trained with CTC, which is much faster than CTC-CRF in retraining. Thus the extra time cost is limited.
As shown in \tabref{table:tab4}, the total time of running the 3-stage NAS procedure is less than 3 times of training a single model from scratch.

\begin{table}[t]
    \centering
    \setlength{\tabcolsep}{3mm}{
        \begin{threeparttable}
            \caption{The estimated run-time for the three stages in the ST-NAS procedure, averaged over the 5 runs in WSJ.}
            \label{table:tab4}
            \begin{tabular}{cccc}
                \toprule
                Stage               & \makecell[c]{warm-up} & \makecell[c]{architecture         \\search} & retraining \\
                \midrule
                Epochs              & 65.2                  & 22                        & 28.6  \\
                Minutes/epoch       & 11                    & 31                        & 27.2  \\
                \midrule
                Total time (minute) & 717.2                 & 682                       & 777.9 \\
                \bottomrule
            \end{tabular}
        \end{threeparttable}}
\end{table}
\begin{figure}[t]
    \begin{minipage}[b]{0.48\linewidth}
        \centering
        \includegraphics[width=4.2cm]{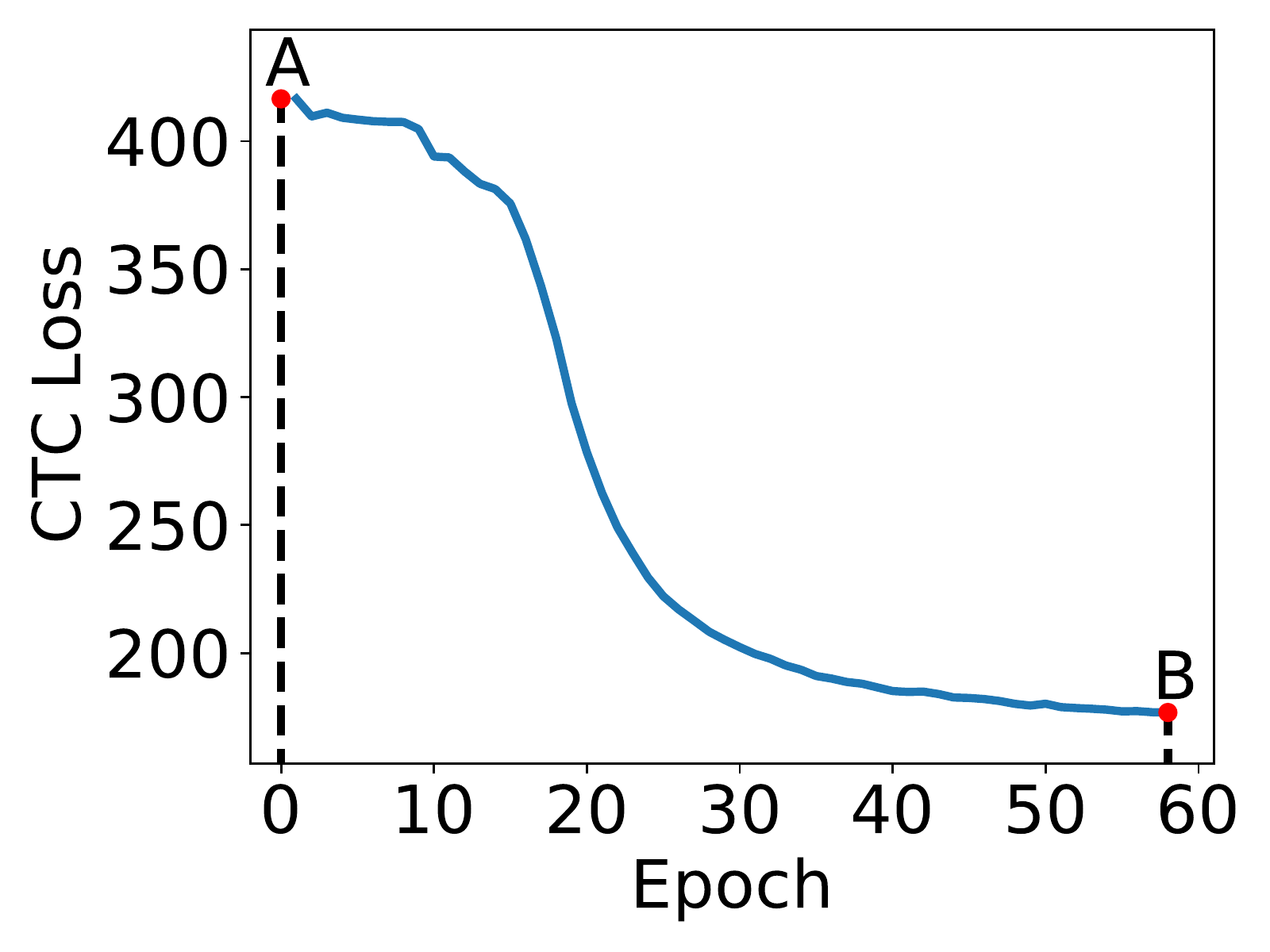}
        \centerline{(a)}
    \end{minipage}
    \begin{minipage}[b]{0.48\linewidth}
        \centering
        \includegraphics[width=4cm]{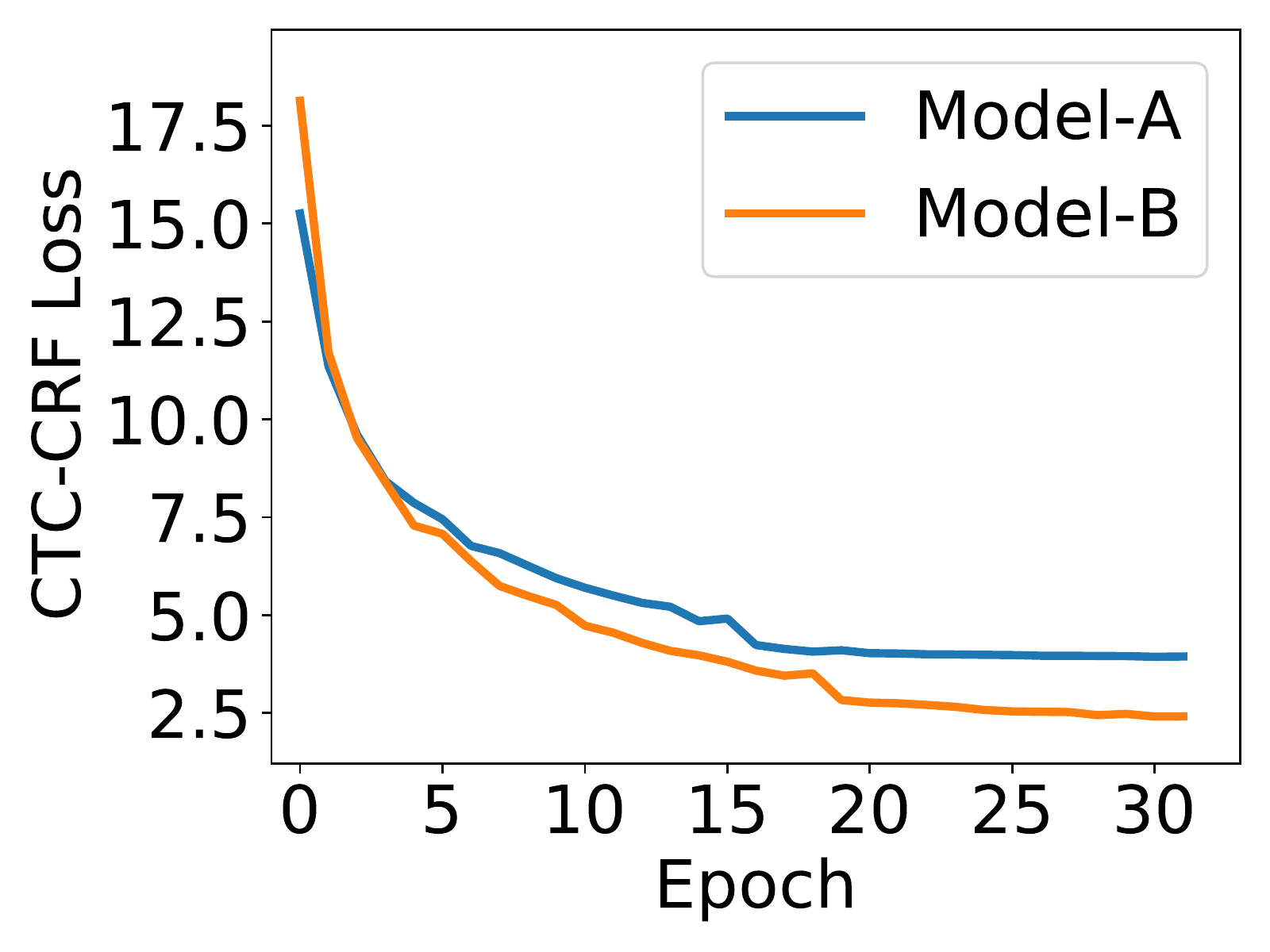}
        \centerline{(b)}
    \end{minipage}
    \vspace{-2mm}
    \caption{(a) The two points A and B in the loss curve during warm-up, which represent differently initialized super-networks. (b) The curves of validation losses for retraining the two models, obtained by running architecture search starting from A and B respectively.}
    \label{fig:method:effectSPI}
\end{figure}

\vspace{-2mm}
\subsection{Effect of super-network initialization}
\label{sec:method:effectSI}
\vspace{-1mm}
The effect of the super-network initialization (i.e., warm-up) seems to be overlooked in previous NAS studies.
We run architecture search from two differently initialized super-networks (A and B), retrain the two searched models, and compare the performance of the two retrained models, as shown in \figref{fig:method:effectSPI}.
Experiments are conducted under the same settings as in \secref{sec:exp:wsj} on the WSJ dataset.
Super-network A represents a randomly-initialized super-network, without any warm-up training, and super-network B is obtained when the warm-up converges.
It can be seen that sufficient warm-up helps the architecture search stage to find a better final model, which achieves lower validation loss in retraining.

\section{Conclusion}
\label{sec:conclusion}
NAS is an appealing next step to advancing end-to-end ASR.
In this paper, we review existing gradient-based NAS methods and develop an efficient NAS method via Straight-Through (ST) gradients, called ST-NAS.
Basically, ST-NAS uses the loss from SNAS but optimizes the loss using the ST gradients.
We successfully apply ST-NAS to end-to-end ASR.
Experiments over WSJ and Switchboard show that the ST-NAS induced architectures significantly outperform the human-designed architecture across the two datasets. Strengths of ST-NAS such as architecture transferability and low computation cost in memory and time are also reported.
Remarkably, the ST-NAS method is flexible and can be further explored by using different macro-DAGs, candidate operations and super-networks for ASR, not limited to the example setup in this paper.

\bibliographystyle{IEEEbib}
\bibliography{refs}

\end{document}